\documentclass[fleqn,twoside,twocolumn,nofootinbib,showkeys]{revtex4} 
\usepackage[nocpr]{ujp} 

\begin{document}
\title[Features of Ultrasound Absorption by Dislocations]
{FEATURES OF ULTRASOUND\\ ABSORPTION BY DISLOCATIONS\\ IN
SUBGRAIN-FREE
Cd\boldmath$_{0.2} $Hg$_{0.8}$Te CRYSTALS}
\author{I.O. Lysyuk}
\affiliation{V.E.~Lashkaryov Institute of Semiconductor Physics,
Nat. Acad.
of Sci. of Ukraine}
\address{41, Nauky Ave., Kyiv 03028, Ukraine}
\email{jaroluk3@ukr.net}
\author{Ya.M.~Olikh}%
\affiliation{V.E.~Lashkaryov Institute of Semiconductor Physics,
Nat. Acad.
of Sci. of Ukraine}%
\address{41, Nauky Ave., Kyiv 03028, Ukraine}%
\author{O.Ya.~Olikh}
\affiliation{Taras Shevchenko National University of Kyiv, Faculty of Physics}
\address{64, Volodymyrs'ka Str., Kyiv 01601, Ukraine}
\email{olikh@univ.kiev.ua}%
\author{G.V.~Beketov\,}%
\affiliation{V.E.~Lashkaryov Institute of Semiconductor Physics,
Nat. Acad.
of Sci. of Ukraine}%
\address{41, Nauky Ave., Kyiv 03028, Ukraine}%
\email{jaroluk3@ukr.net} \udk{???} \pacs{62.65.+k} \razd{\secvii}

\autorcol{I.O.\hspace*{0.7mm}Lysyuk, Ya.M.\hspace*{0.7mm}Olikh,
O.Ya.\hspace*{0.7mm}Olikh et al.}

\setcounter{page}{50}%

\begin{abstract}
The temperature dependence of the ultrasound wave absorption in bulk
$p$-Cd$_{0.2}$Hg$_{0.8}$Te crystals free from low-angle grain
boundaries has been studied experimentally for the first time in the
frequency range 10--55~MHz and the temperature interval 150--300~K,
and the corresponding results of measurements are presented. The
maximum value of absorption coefficient is found to increase and
to shift toward higher temperatures, as the ultrasound frequency grows.
The results obtained can be satisfactorily explained in the
framework of the Brailsford model, which associates the ultrasound
absorption with vibrations of thermally activated dislocation kinks.
The characteristic parameters of this model for
$p$-Cd$_{0.2}$Hg$_{0.8}$Te are determined; namely, the frequency
coefficient $f_{k}\approx $~6\,$\times $\,10$^{\,9}$\,Hz and the
kink diffusion activation energy $W_{k}\approx $~0.11\,eV. The
dislocation concentration is also evaluated ($\alpha \approx
$~2\,$\times$\,10$^{10}$\,m$^{-2}$), with the determined value being
consistent with that obtained by the selective etching method
(0.7\,$\times $\,10$^{10}$\,m$^{-2}$).
\end{abstract}
\keywords{ultrasound, dislocations, Cd$_{x}$Hg$_{1-x}$Te.}
\maketitle

\section{Introduction}

The more than semicentenial history of researches and the
implementation into practice of photo-sensitive CdHgTe crystals
showed that they remain a material of choice for IR radiation
detectors \cite{Ponom,Gasan}. The basic electrophysical and
photo-electric properties of this material are governed by extremely
high concentrations of electrically active point-like
($10^{21}$--$10^{22}$~m$^{-3}$) and linear (a density of
$10^{9}$--$10^{10}$~m$^{-2}$ for the growth dislocations) defects
which interact with one another. The application of ultrasound is
one of the methods to controllably vary the defect structure in both
CdHgTe and other semiconductors \cite{Ostrov,OlOl,OOl2}. In
particular, it was found that the acoustically stimulated
reconstruction of point defects, owing to their electric and
deformation interaction with dislocations, results in changes in the
concentration, mobility, and lifetime of free charge carriers in
CdHgTe \cite{OlShav,Anna}. However, for obtaining predictable
results of such an influence, we must know the mechanism of
interaction between ultrasonic waves and the crystal. The previous
acoustic researches for CdHgTe \cite{OlSal,AnnaSal,Kalit,OlDub} were
carried out with the use of lamellar specimens possessing a subgrain
structure in the wafer plane. In particular, with the help of the
internal friction (IF) method, it was found that the resonance
absorption at low-angle boundaries (LABs) and the dislocation
absorption in bulk are the dominating mechanisms of ultrasound
losses in such specimens, with the latter process being well
described with the use of the Granato--Lucke string model
\cite{Granato}. The described character of amplitude-dependent
changes in the IF method is known to be realized at large amplitudes
of mechanical stresses, when dislocations can detach from the weak
pinning centers formed by point defects \cite{Nik}.

At the same time, provided that there exist the dislocation loops with short
lengths (stoppers with weak pinning), and the mechanical deformations are
small, another type of acoustic dislocation absorptions has to take place,
when losses have a relaxation character and depend on the ultrasound
frequency \cite{Blist,True}. Such a process of amplitude-independent
absorption can be realized in the case where low-intensity ultrasound waves
are used, e.g., in the pulse mode. This mode was used to measure the elastic
moduli in CdHgTe \cite{Vasil,Machulin}. However, there are almost no results
obtained for the temperature dependence of the absorption coefficient for
ultrasound waves in this material. This circumstance is related to
complicated conditions imposed on the geometry and the structural quality of
specimens intended for acoustic measurements \cite{True}, as well as to
the technological complexity of manufacturing the crystallographically oriented
subgrain-free CdHgTe specimens.

Note that CdHgTe is an interesting model material, in view of the
features of its crystalline structure, to study dislocations with
the use of the ultrasound technique. Really, the motion of
dislocations in a semiconductor occurs in a medium with a high
concentration of electrically active point defects, which are
closely connected with dislocations and essentially affect their
motion.

This work was aimed at the experimental study of the temperature
dependences of the ultrasound absorption coefficient in
subgrain-free Cd$_{0.2}$Hg$_{0.8}$Te crystals and at the search of
adequate models for the description of the acousto-dislocation
interaction in them.\vspace*{-2mm}

\section{Specimens}

Single-crystalline specimens $10\times 6\times 2.5~\mathrm{mm}^{3}$
in size and oriented in the directions $\left\langle
100\right\rangle $ and $\left\langle 110\right\rangle $ with an
accuracy of $2^{\circ }$ were fabricated from an
$p$-Cd$_{0.2}$Hg$_{0.8}$Te ingot grown up at the affiliate
\textquotedblleft Pure Metal Plant\textquotedblright\ of the public
corporation \textquotedblright Pure Metals\textquotedblright\
(Svitlovodsk). The planeness and the parallelism of lateral edges
satisfied the condition ${\Delta s}/{s}<10^{-4}$, where $s$ is the
specimen length, and ${\Delta s}$ its variation at various points,
which is necessary for the measurements of acoustic characteristics
at multiple reflections of an ultrasound wave to be correct.

The high structural quality of examined specimens was verified by the
reliable x-ray orientation of two crystallographic planes. Note that the
experimental specimens were used earlier to study the anisotropy of CdHgTe
elastic properties \cite{Machulin}, and the results obtained agreed well
with those of work \cite{Vasil}.

\begin{figure}%
\vskip1mm
\includegraphics[width=\column]{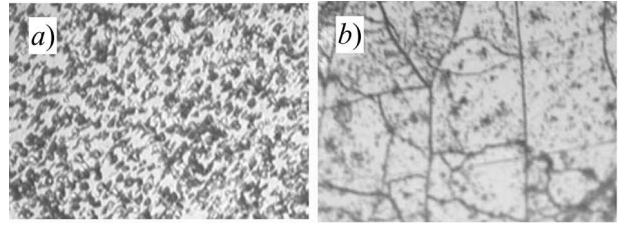}
\vskip-3mm\caption{Photo of the selectively etched surface of
studied Cd$_{0.2}$Hg$_{0.8}$Te specimens (\textit{a})~without and
(\textit{b})~with a subgrain structure. Panel (\textit{b}) was taken
from work \cite{Olikh}  }
\end{figure}

To study the dislocation structure in Cd$_{0.2}$Hg$_{0.8}$Te wafers,
the method of dislocation etch was used. Immediately before the
etching procedure, the specimen surface was trea\-ted with a
polishing solution (0.2~ml of Br$_{2}$~+ 10~ml of CH$_{3}$OH). Then
the specimen was carefully washed out in, first, ethylene glycol
and, afterward, isopropanol. For the selective etching, which was
carried out at room temperature, the Schaake etchant (5~g of
CrO$_{3}$~+ 3~ml of HCl~+ 15~ml of H$_{2}$O) was used, which allowed
dislocations to be revealed on both polar edges (111)A and (111)B,
as well as on the surfaces with intermediate orientations
\cite{Capp}. The etching time varied from 20 to 60~s for various
specimens. In Fig.~1,~\textit{a}, a typical photo of the researched
specimen surface after its selective etching is shown. The obtained
picture testifies to a uniform distribution of dislocations and no
LAB formation. For comparison, a photo of a typical etched surface
of Cd$_{0.2}$Hg$_{0.8}$Te crystal that contains LABs \cite{Olikh} is
exhibited in Fig.~1,~\textit{b}. The evaluated dislocation
concentration showed that, according to the selective etching data,
$\Lambda _{\rm SE}=(6\div 8)\times 10^{9}$~m$^{-2}$ in the studied
specimens.

\section{Measurement Procedure}

In order to measure the ultrasound absorption coefficient $\alpha $,
the pulse-echo method was used. The block diagram of the
corresponding experimental installation is shown in Fig.~2.
Ultrasound pulses in the specimens were excited with the help of a
piezoelectric transducer, and LiNbO$_{3}$ wafers with the
$(Y+36^{\circ })$-cut were used for this purpose. The generated
train of radio pulses was directed onto a piezoelectric transducer.
The acoustic contact was provided in a wide temperature interval by
applying silicone oil of the GKZh-94 type. Ultrasound pulses were
reflected multiple times from the specimen end faces, so that a
series of echoed radio pulses was formed on a piezoelectric
receiver, which was identical to the emitting transducer. Video
signals were monitored on an oscilloscope screen and,
simultaneously, were directed to a stroboscopic converter and,
afterward, to a personal computer. With the help of the developed
software, the amplitudes of video signals were measured and averaged
in time. We also monitored the time delay between the video-signal
maxima and the synchronization pulse, which allowed us to account
for the effect of temperature-induced variation of the ultrasound
velocity.

\begin{figure}%
\vskip1mm
\includegraphics[width=7cm]{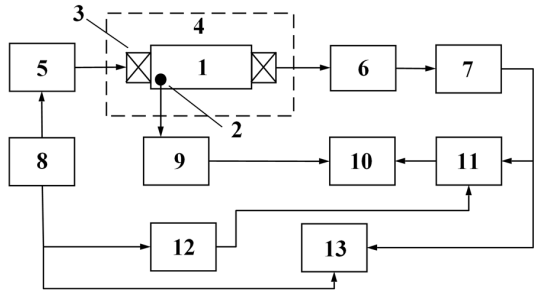}
\vskip-2mm\caption{Block diagram of the installation to measure the
ultrasound wave absorption coefficient in solid-state specimens with
the help of the pulse-echo technique: ({\it 1})~specimen, ({\it
2})~thermocouple, ({\it 3}) piezoelectric transducer, ({\it
4})~cryostat, ({\it 5})~high-frequency generator G4-151, ({\it
6})~high-frequency amplifier U3-28, ({\it 7})~amplitude detector,
({\it 8})~square-wave generator G5-54, ({\it 9})~digital voltmeter
V7-21A, ({\it 10})~computer, ({\it 11})~stroboscopic converter V9-5,
({\it 12})~generator of delay pulses, and ({\it 13})~oscilloscope
S1-98 }
\end{figure}

\begin{figure}%
\vskip4mm
\includegraphics[width=\column]{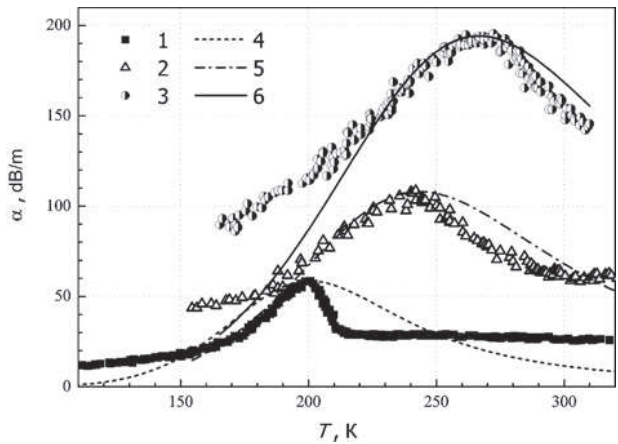}
\vskip-3mm\caption{Experimental (symbols) and theoretical (curves,
Eq.~(\ref{e12})) temperature dependences of the absorption
coefficient for longitudinal ultrasonic waves with various
frequencies $f_{\mathrm{us}}$ at their propagation along the
direction $\left\langle 110\right\rangle $\thinspace in the
$p$-Cd$_{0,2}$Hg$_{0,8}$Te crystal: $f_{\mathrm{us}}=11.2$
(\textit{1} and \textit{4}), 34.8 (\textit{2} and \textit{5}), and
55.4~MHz (\textit{3} and \textit{6}). The parameters used in
calculations are $W_{k}=0.108$~eV, $f_{k}=6.03\times10^{9}$~Hz, and
$g=3.5$  }\vspace*{2mm}
\end{figure}

The application of pulse excitation and high-frequency low-amplitude
signals was aimed at implementing the mode of amplitude-independent
ultrasound absorption. Our measurements confirmed that the value of
$\alpha$ really remains constant at room temperature and at the
exciting radio pulse amplitude \mbox{up to 3~V.}

The absolute values of ultrasound absorption were measured to an accuracy of
10\%.

\section{Experimental Results and Their Discussion}

In Fig.~3, the experimentally measured temperature dependences
of the absorption coefficient $\alpha $ for longitudinal ultrasound
waves with various frequencies propagating in
$p$-Cd$_{0.2}$Hg$_{0.8}$Te crystals along the direction
$\left\langle 110\right\rangle $ are exhibited. One can see that
each $\alpha (T)$-dependence has a maximum, and the temperature of
the maximum, $T_{m}$, and the maximum absorption value $\alpha _{m}$
depend on the ultrasound frequency $f_{\mathrm{us}}$. In particular,
when $f_{\mathrm{us}}$ increases from 11 to 55~MHz, the value of
$\alpha _{m}$ grows from 60 to 190~dB/m and that of $T_{m}$ from 199
to 270$~\mathrm{K}$ (see also Table).

\begin{table*}[!]
\vskip4mm \noindent\caption{Ultrasound absorption parameters for
\boldmath$p$-Cd$_{0.2}$Hg$_{0.8}$Te
crystals}\vskip3mm\tabcolsep10.0pt

\noindent{\footnotesize
\begin{tabular}{|c|c|c|c|c|c|c|c|c|}
 \hline%
\raisebox{-7mm}[0cm][0cm]{$f_{\rm us}$, MHz}&
\multicolumn{2}{|c}{\raisebox{-3mm}[0cm][0cm]{Experiment}}
 & \multicolumn{6}{|c|}{\rule{0pt}{5mm}Theory}\\[1.5mm]%
\cline{4-9}
 &\multicolumn{2}{c}{}&\multicolumn{3}{|c}{\rule{0pt}{4mm}Granato--Lucke
model \cite{True}}
&\multicolumn{3}{|c|}{Brailsford model \cite{Bra}}\\[1.5mm]
\cline{2-9}

&\rule{0pt}{4mm}$T_m$,&$\alpha_m$,&$\omega_0$,&$B_m$,&$L$,&$W_k$,&$f_k$,&$\Lambda_B$,\\

&K&dB$\cdot$m$^{-1}$&$10^8$ rad$\cdot$с$^{-1}$
 &$10^{-5}$~kg/m/s&$10^{-6}$~m&eV&$10^9$~Hz&$10^{10}$~m$^{-2}$\\[1.5mm]
\hline

\rule{0pt}{4mm}11.2&$200\pm3$&$59\pm5$&6.8&3.3&4.1&&&2.3\\
34.8&$242\pm3$&$108\pm5$&9.1&1.8&3.1&0.108&6.03&1.7\\
55.4&$269\pm3$&$194\pm5$&9.0&1.0&3.1&&&2.1\\[2mm]
\hline
\end{tabular}
}\label{tbl}
\end{table*}

The absorption of ultrasound waves in crystals can be driven by
various mechanisms \cite{True}. However, the estimations testify
that, under the actual experimental conditions (the temperature
interval $80-400$~K and the ultrasound frequency range 1--100~MHz),
the ultrasound damping arising owing to phonon-phonon processes and
thermo-elastic losses can be neglected, because the absorption
coefficient in those cases does not exceed 1 and 10$^{-4}$~dB/m,
respectively \cite{True}. On the other hand, as was already
mentioned above, the literature data indicate that the ultrasound
wave damping in subgrain $p$-Cd$_{x}$Hg$_{1-x}$Te crystals with
$x\approx 0.2$ is mainly governed by the dislocation absorption and
the resonance absorption at LABs \cite{Kalit,OlDub,Olikh}. Hence, a
conclusion can be drawn that the acousto-dislocation interaction is
the main mechanism of ultrasound absorption in subgrain-free
$p$-Cd$_{0.2}$Hg$_{0.8}$Te specimens with high dislocation
concentrations. Proceeding from it, let us analyze the obtained
experimental results in the framework of some known \mbox{models.}

\subsection{Granato--Lucke model}

Despite that the model of dislocation friction by Granato and Lucke
was  developed in the idealizing zero-temperature approximation for
a crystal, it can be successfully applied to analyze the dislocation
internal friction in various real materials, in particular, in
semiconductor crystals \cite{Ostrov,Nik,Blist}. In the framework of
this approach, a dislocation is considered as a string pinned at
definite points; free string sections between pinning points can
vibrate under the action of an external force, in particular,
ultrasound. In this case, the coefficient of acoustic wave
absorption at low frequencies, $\omega =2\pi f_{\mathrm{us}}\ll
\omega _{0}$, where $\omega _{0}$ is the characteristic frequency of
vibrations for the dislocation section, has to be described by the
following relation \cite{Granato,True}:
\begin{equation}
\alpha =\frac{4G\Lambda \omega ^{2}}{\upsilon \pi ^{3}\rho }
\frac{d}{(\omega _{0}^{2}-\omega ^{2})^{2}+d^{2}\omega ^{2}},
\label{e1}
\end{equation}%
where $G$ is the shear modulus, $\upsilon $ the velocity of ultrasound wave
propagation, $\rho $ the material density, $\omega _{0}^{2}=2G/[L^{2}(1-\mu
)\rho ]$, $L$ the length of dislocation section; $\mu $ Poisson's ratio,
$d=B/(\pi \rho b^{2})$ the damping constant, $B$ the coefficient of
dynamic viscosity, and $b$ the absolute value of Burgers vector. Let us try
to estimate the individual values of parameters from the obtained
experimental data.

It is known from the literature that the variations of the elastic moduli for
Cd$_{x}$Hg$_{1-x}$Te crystals do not exceed $1-2\%$ in the experimental
temperature interval \cite{Machulin}. If we suppose that $L$ and $\rho $
also weakly depend on the temperature, then, in accordance with
Eq.~(\ref{e1}), the temperature dependence of the ultrasound wave absorption coefficient must
be associated with variations of the parameter $d$, i.e., actually, with
the coefficient of dynamic viscosity. The results of calculations showed
that the parameter $\alpha $ must reach its maximum value at $d_{m}=(\omega
_{0}^{2}-\omega ^{2})/\omega $, and the following relations are valid at
this point:
\begin{equation}
\alpha _{m}=\frac{2G\Lambda \omega }{\upsilon \pi ^{3}\rho (\omega
_{0}^{2}-\omega ^{2})},  \label{e2}
\end{equation}\vspace*{-5mm}
\begin{equation}
\omega _{0}=\sqrt{\frac{2G\Lambda \omega }{\upsilon \pi ^{3}\rho
\alpha _{m}}+\omega ^{2}}\approx \sqrt{\frac{2G\Lambda \omega
}{\upsilon \pi ^{3}\rho \alpha _{m}}},  \label{e3}
\end{equation}
\begin{equation}
B_{m}=\frac{2G\Lambda b^{2}}{\upsilon \pi ^{2}\alpha _{m}},
\label{e4}
\end{equation}\vspace*{-5mm}
\begin{equation}
L=\sqrt{\frac{2G}{(1-\mu )\rho \omega _{0}^{2}}},  \label{e5}
\end{equation}%
where $B_{m}$ is the value of coefficient $B$ when the absorption is maximum.

The individual values of parameters, which were obtained with the
help of expressions (\ref{e3})--(\ref{e5}) and with the use of
experimentally determined values for $\alpha_{m}$, are given in
Table. In the calculations, the following parameter values were
selected for Cd$_{0.2}$Hg$_{0.8}$Te: $b=$
$=4.58\times10^{-10}$~m~\cite{Morgan}, $\rho=7.625\times$ $\times
10^{3}$~kg$/$m$^{3}$, $\mu=$ $=0.365$, $G=1.95\times10^{10}$~Pa,
$\upsilon=3.0\times10^{3}$~m/s~\cite{Erof}, and
$\Lambda=\Lambda_{\rm SE}=7\times10^{9}$~m$^{-2}$.

Note that, as a result of the use of the Granato--Lucke model
to explain the appearance of a maximum in the dependence $\alpha
(T)$, we obtained that, in accordance with
Eqs.~(\ref{e2})--(\ref{e5}), the value of $\alpha _{m}$ has to be
proportional to the ultrasound frequency, because the resonance
vibration frequency on the dislocation section and its length have to
be independent of the external perturbation frequency. In general,
the experimentally obtained data agree with this claim
(Fig.~4,~\textit{a}). Moreover, if the dislocation grid nodes are
supposed to be the main points of the linear-defect pinning, the
corresponding calculated value $L\approx 3.5\times 10^{-6}$~m also
satisfactorily correlates with the average distance between
dislocation lines, $R=1/\sqrt{\Lambda _{\rm SE}}\approx 1\times
10^{-5}$~m.

\begin{figure*}%
\vskip1mm
\includegraphics[width=16.7cm]{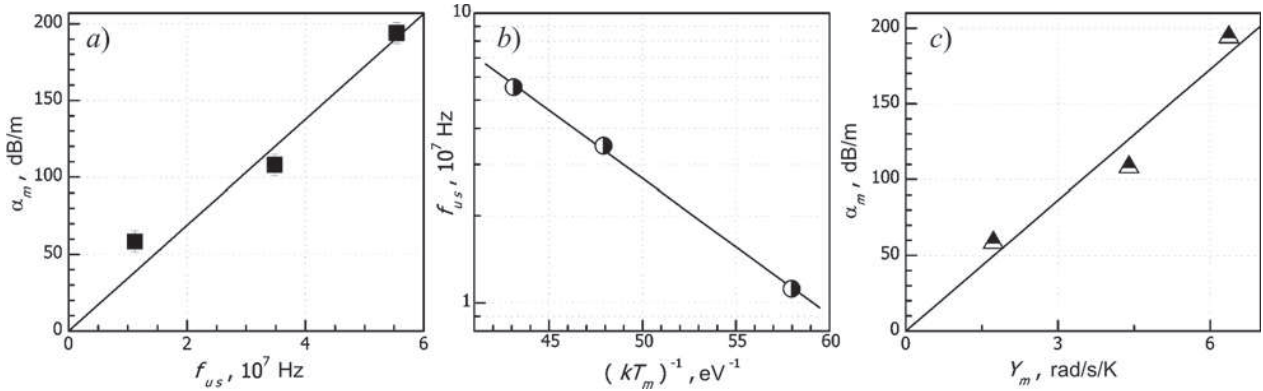}
\vskip-3mm\caption{Dependences of (\textit{a})~the value of
absorption coefficient maximum on the ultrasound frequency
$f_{\mathrm{us}}$, ({\it b})~the value of $f_{\mathrm{us}}$ for the
maximum ultrasound absorption on the reciprocal temperature, and
({\it c})~the maximum value of ultrasound absorption coefficient at
a definite frequency on the value of function $Y$ at its maximum.
While calculating $Y_{m}$, the following parameter values were used:
$W_{k}=0.108$~eV, $f_{k}=6.03\times 10^{9}$~Hz, and $g=3.5$  }
\end{figure*}

On the other hand, this model predicts that hindering the
dislocation motion, including the vibrational one, in the ultrasound
field occurs owing to the interaction of dislocations with phonons
and charge carriers, and because of thermoelastic losses
\cite{Granato,Sudz,True}. The temperature dependence of each of
those mechanisms, as well as their relative contributions to the
$B$- and $d$-values, can substantially depend on the specific
material. Therefore, it is rather difficult to describe the
dependence $\alpha (T)$ precisely in the framework of the
Granato--Lucke model. However, it should be taken into account that,
as was shown in many works (in particular, in \cite{Nik,True}), the
value of $B$ linearly increases in the high-temperature interval
($T>\Theta _{D}$, where $\Theta _{D}$ is the Debye temperature) with
the temperature. For Cd$_{x}$Hg$_{1-x}$Te, the Debye temperature
$\Theta _{D}\approx 146$~K \cite{Agaev}. However, the $B_{m}$-values
obtained at various $T_{m}$ contradict the growing behavior expected
for the temperature dependence. Therefore, the Granato--Lucke model
cannot be used to quantitatively analyze the temperature dependence
of the ultrasound absorption coefficient for the given
\mbox{material.}

\subsection{Brailsford model}

This model was proposed by Brailsford \cite{Bra} in order to explain
the characteristic peaks of the ultrasound-wave absorption that were
observed in plastically deformed metals at low temperatures.
According to this model, a dislocation is regarded as a sequence of
segments oriented in the dense packing direction and connected by
abrupt kinks. The dislocation is considered to be rigidly pinned at
the end points, and ultrasound is absorbed due to the stimulated
motion of kinks. It is supposed that the diffusion of kinks has a
thermally activated character, and the diffusion coefficient $D$ is
described by the expression $D=D_{0}\exp {(-W_{k}/kT)}$, where
$W_{k}$ is the diffusion activation energy. In work \cite{Bra}, it
was also shown that the relation between the ultrasound wave
frequency $f_{\mathrm{us}}$ and the absorption maximum temperature
$T_{m}$ can be expressed by the formula
\begin{equation}
f_{\rm us}=f_{k}\exp {\left(\! -\frac{W_{k}}{kT_{m}}\!\right) }\!,
\label{e6}
\end{equation}%

\noindent where $f_{k}=\pi D_{0}/(20\,l_{0}^{2})$, and $l_{0}$ is
the average length of dislocation segments. As is seen from
Fig.~4,~\textit{b}, the experimental dependence of the ultrasound
frequency on the reciprocal temperature of the absorption maximum is
really a straight line on the semilogarithmic scale. By calculating
a linear approximation of the data depicted in Fig.~4,~{\it b}, we
obtained the following values for the used
$p$-Cd$_{0.2}$Hg$_{0.8}$Te specimens: $W_{k}=0.108\pm 0.004$~eV and
$f_{k}=(6.03\pm 0.05)\times 10^{9}$~Hz.

According to work \cite{Bra}, the Q-factor $Q_{l}$ corresponding to
the absorption of ultrasound waves by a single dislocation segment
of length $l$ is described by the expression
\begin{equation}
\label{e7} Q_l^{-1}\!=\!\frac{8Ga^2b^2l^3(n_0\!+\!p_0)}{VkT\pi^4}
\frac{\left(\!\frac{\omega l^2}{20\pi
l_0^2f_k}\!\right)\exp{\left(\!\frac{W_k}{kT}\!\right)}}
{1\!+\!\left(\!\frac{\omega l^2}{20\pi
l_0^2f_k}\!\right)^{\!\!2}\!\exp{\left(\!\frac{2W_k}{kT}\!\right)}}\!,
\end{equation}
where $a$ is the lattice constant ($a=6.466\times10^{-10}$~m
\cite{Morgan}); $n_{0}$ and $p_{0}$ are the equilibrium linear
concentrations of the right and left kinks, respectively; and $V$ is
the crystal volume. The total number of segments, $N$, is connected
with the dislocation concentration as follows:
\begin{equation}
\Lambda=\frac{Nl_{0}}{V}.  \label{e8}
\end{equation}

The theory says that, in order to estimate the general $Q$ factor of
a crystal, it is necessary to multiply $N$ by the averaged value
of the quantity described by expression (\ref{e7}), and the
averaging must be carried out with regard for the distribution
of segments over their lengths. If we assume that
the averaging is equivalent to the substitution of the parameter $l$ in
Eq.~(\ref{e7}) by a definite effective segment length
$l_{\mathrm{eff}}=gl_{0}$ and make allowance for the relation
between $\alpha $ and $Q$--namely, $Q^{-1}=\alpha \upsilon \ln
{10}/(10\,\omega )$--then the expression describing the absorption
according to the Brailsford model reads
\begin{equation} \alpha
(T,\omega )=\frac{4Ga^{2}b^{2}g^{3}D_{0}(n_{0}+p_{0})\Lambda }{\ln
{10}\,\upsilon f_{k}k\pi ^{3}} Y(T,\omega ),  \label{e9}
\end{equation}where the function
\begin{equation}
\label{e10} Y(T,\omega)=\frac{\omega}{T}\frac{\left(\frac{\omega
g^2}{20\pi f_k}\right)\exp{\left(\!\frac{W_k}{kT}\!\right)}}
{1+\left(\!\frac{\omega g^2}{20\pi
f_k}\!\right)^{\!\!2}\exp{\left(\!\frac{2W_k}{kT}\!\right)}}
\end{equation}
mainly governs the temperature and frequency dependences of
the absorption coefficient. The function $Y(T,\omega )$ is plotted in
Fig.~5. One can see that its maximum shifts toward higher
temperatures, as the frequency increases. Just this behavior was
observed experimentally for the parameter $\alpha $ (see Table). The
agreement between the experimental $T_{m}$-values and the maximum
positions of the function $Y(T,\omega )$ at the ultrasound frequencies
that were used in the experiments is reached at $g=3.5.$

Notice that, in accordance with Eq.~(\ref{e9}), we obtain
\begin{equation}
\alpha _{m}=\frac{4Ga^{2}b^{2}g^{3}D_{0}(n_{0}+p_{0})\Lambda }{\ln
{10}\upsilon f_{k}k\pi ^{3}} Y_{m},  \label{e11}
\end{equation}%
where $Y_{m}=Y(T_{m},\omega )$, i.e. the quantities $\alpha _{m}$
and $Y_{m}$ must be proportional to each other, which was really
obeyed in the experiment with a rather high accuracy (see
Fig.~4,~\textit{c}). At the same time, expression (\ref{e9}) can be
rewritten in the form
\begin{equation}
\alpha (T,\omega )=\frac{\alpha _{m}}{Y_{m}} Y(T,\omega ).
\label{e12}
\end{equation}%
In Fig.~3, the results of calculations according to relation (\ref{e12}) are
shown. One can see that they describe the experimental dependences rather well
qualitatively and, at high frequencies, even quantitatively. Hence, the
results obtained testify to the expediency of the Brailsford model
application to the description of ultrasound absorption processes in
subgrain-free Cd$_{x}$Hg$_{1-x}$Te crystals.

\begin{figure}%
\vskip1mm
\includegraphics[width=\column]{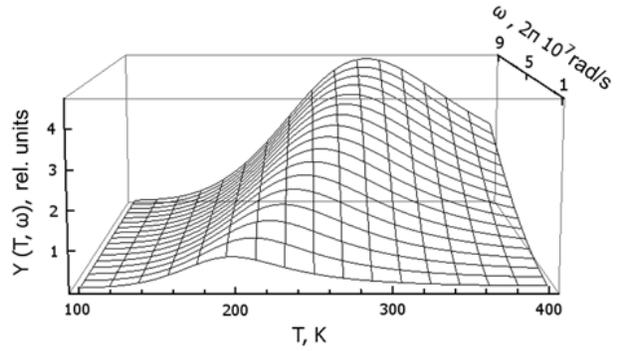}
\vskip-3mm\caption{Temperature-frequency dependence of the function
$Y(T,\omega)$. The parameters of calculations by formula (\ref{e10})
are $W_{k}=0.108$~eV, $f_{k}=6.03\times10^{9}$~Hz, and $g=3.5$  }
\end{figure}

The Brailsford model makes it possible to use the experimental data
obtained for the absorption of ultrasound waves to use for the
estimation of the dislocation concentration $\Lambda _{B}$ as well.
Really, as was shown in work \cite{Bra}, $D_{0}\propto a^{2}\nu
_{D}$ and $(n_{0}+p_{0})\approx \left\vert \tan {\theta }\right\vert
a_{s}^{-1}$, where $\nu _{D}$ is the Debye frequency ($\nu
_{D}=3\times 10^{12}$~Hz for Cd$_{0.2}$Hg$_{0.8}$Te \cite{Agaev}),
$a_{s} $ is the interplane distance, and $\theta $ the angle between
the dense packing direction and the dislocation line ($\theta \leq
\pi /6$). If we take the values $\theta =\pi /12$ and
$a_{s}=a/\sqrt{3}$ (this is the value for the $\left\langle
110\right\rangle $ direction \cite{Hirt}), we obtain from
Eq.~(\ref{e11}) that
\begin{equation}
\Lambda _{B}\propto \frac{1.24\,\upsilon f_{k}k\,\pi ^{3}\alpha
_{m}}{Ga^{3}b^{2}g^{3}\nu _{D}Y_{m}}.  \label{e13}
\end{equation}

The values of $\Lambda _{B}$ calculated by formula (\ref{e13}) for
three ultrasound frequencies are listed in Table. We would like to
emphasize a satisfactory agreement between these values and the
$\Lambda _{\rm SE}$-value obtained in the framework of the selective
etching method. The corresponding difference can be explained as
follows: first, the relation $D_{0}\propto a^{2}\nu _{D}$ is
approximate, and, second, the estimation of a dislocation concentration
using the selective etching method gives, as a rule, a somewhat
underestimated value \cite{Ovsuk}. Moreover, we should mark a
definite shortcoming of such an analysis. It is associated with the
application of rather a symbolic value ($0.46a^{-1}$) for the total
kink number, although this parameter considerably depends on the
orientation of the dislocation line with respect to the Peierls valley.

Note that the model describing the motion of elements in the fine
dislocation structure was successfully developed by Loktev and Khalack
\cite{Loktev} to explain the amplitude-dependent effects under the action of an
intensive ultrasound wave, in particular the sonoluminescence effect in CdS.

\section{Conclusions}

In this work, the temperature dependences of the coefficient of
longitudinal bulk acoustic wave absorption, $\alpha$, in bulk
$p$-Cd$_{0.2}$Hg$_{0.8}$Te specimens free of low-angle boundaries
were experimentally studied for the first time within the
temperature interval 150--300$~\mathrm{K}$ and the frequency range
10--55~MHz. The measured dependences turned out nonmonotonous. As
the ultrasound frequency increases, both the absorption maximum and
the corresponding tempera- \mbox{ture grow.}\looseness=1

The data obtained were analyzed in the framework of the classical
Granato--Lucke model, which allowed us to estimate the length of
dislocation sections ($L\approx 3.5\times 10^{-6}$~m) and the
characteristic frequency of their vibrations ($\omega _{0}\approx
9\times 10^{8}$~rad/s). It was shown that the temperature dependence
of $\alpha $ for subgrain-free $p$-Cd$_{0.2}$Hg$_{0.8}$Te crystals
can be explained using the Brailsford model, which associates the
absorption of ultrasound waves with the motion of thermally
activated dislocation kinks. In the framework of this model, the
activation energy of the kink diffusion, $W_{k}\approx 0.11$~eV, and
the frequency parameter, $f_{k}\approx 6\times 10^{9}$~Hz, were
determined. The Brailsford model was also used to estimate the
dislocation concentration by analyzing the dependence $\alpha (T)$.
The corresponding estimated value ($2\times 10^{10}$~m$^{2}$) is
shown to agree well with that obtained in the framework of the
selective etching method \mbox{($0.7\times
10^{10}$~m$^{2}$).}\looseness=1

\vskip4mm{\it  The authors express their gratitude to
K.P.\,Kur\-ba\-nov for the fabrication of bulk single-grain
Cd$_{0.2}$Hg$_{0.8}$Te specimens.}

\rezume{%
І.О. Лисюк, Я.М. Оліх, О.Я. Оліх, Г.В. Бекетов}{ОСОБЛИВОСТІ
ДИСЛОКАЦІЙНОГО\\ ПОГЛИНАННЯ УЛЬТРАЗВУКУ В БЕЗСУББЛОЧНИХ\\ КРИСТАЛАХ
Cd$_{0,2}$Hg$_{0,8}$Te\vspace*{-1mm}} {Вперше наведено результати
експериментального дослідження температурної (150--300~K) залежності
поглинання ультразвукових (УЗ) хвиль у об'ємних кристалах
$p$-Cd$_{0{,}2}$Hg$_{0{,}8}$Te, які не містять малокутових границь,
у частотному діапазоні 10--55~МГц. Виявлено, що при збільшенні
частоти УЗ спостерігається збільшення максимального значення
коефіцієнта поглинання та його зсув у бік високих температур.
Показано, що експериментальні результати задовільно пояснюються в
рамках моделі Брейсфолда, яка розглядає поглинання УЗ хвилі за
рахунок коливання термоактивованих дислокаційних перегинів.
Визначено характерні параметри даної моделі для
$p$-Cd$_{0,2}$Hg$_{0,8}$Te, а саме частотний фактор
$(6\cdot10^9$~Гц) та активаційна енергія руху перегинів
$(0{,}11$~еВ), а також проведено оцінку густини дислокацій
$(2\cdot10^{10}$~м$^{-2})$, яка узгоджується з даними, отриманими
методом селективного травлення $(0{,}7\cdot10^{10}$~м$^{-2})$.}

\end{document}